\documentclass[twocolumn,prb,showpacs,amssymb,amsmath,superscriptaddress]{revtex4-1}
\usepackage{hyperref}
\usepackage[latin9]{inputenc}
\usepackage{color}
\usepackage{verbatim}
\usepackage{amsmath}
\usepackage{graphicx}

\makeatletter

\usepackage{bm}
\def\be{\begin{align}}
\def\ee{\end{align}}
\def\bea{\begin{eqnarray}}
\def\eea{\end{eqnarray}}

\newcommand{\ket}[1]{|#1\rangle}
\newcommand{\bra}[1]{\langle#1|}

\newcommand{\sgn}{\mathrm{sgn}}

\providecommand{\abs}[1]{\vert#1\vert}
\providecommand{\e}{\text{e}} 

\makeatother

\begin{document}

\title{Effect of interactions on quantum-limited detectors}

\author{Gleb Skorobagatko}

\affiliation{%
\mbox{%
Dahlem Center for Complex Quantum Systems and Fachbereich Physik,
Freie Universit\"at Berlin, 14195 Berlin, Germany%
}}

\author{Anton Bruch}

\affiliation{%
\mbox{%
Dahlem Center for Complex Quantum Systems and Fachbereich Physik,
Freie Universit\"at Berlin, 14195 Berlin, Germany%
}}

\author{Silvia {Viola Kusminskiy}}

\affiliation{%
\mbox{%
Institute for Theoretical Physics, University Erlangen-N\"urnberg,
Staudtstra\ss{}e 7, 91058 Erlangen, Germany%
}}

\affiliation{%
\mbox{%
Dahlem Center for Complex Quantum Systems and Fachbereich Physik,
Freie Universit\"at Berlin, 14195 Berlin, Germany%
}}

\author{Alessandro Romito}

\affiliation{Department of Physics, Lancaster University, Lancaster LA1 4YB, United
Kingdom}

\affiliation{%
\mbox{%
Dahlem Center for Complex Quantum Systems and Fachbereich Physik,
Freie Universit\"at Berlin, 14195 Berlin, Germany%
}}
\begin{abstract}
We consider the effect of electron-electron interactions
on a voltage biased quantum point contact in the tunneling regime
used as a detector of a nearby qubit. We model the leads of the quantum
point contact as Luttinger liquids, incorporate the effects of finite
temperature and analyze the detection-induced decoherence
rate and the detector efficiency, $Q$. We find that interactions
generically reduce the induced decoherence along with the detector's
efficiency, and strongly affect the relative strength of the decoherence
induced by tunneling and that induced by interactions with the local density. With
increasing interaction strength, the regime of quantum-limited detection
($Q \to 1$) is shifted to increasingly lower temperatures or higher bias
voltages respectively. For small to moderate interaction strengths,
$Q$ is a monotonously decreasing function of temperature as in the
non-interacting case. Surprisingly, for sufficiently strong interactions
we identify an intermediate temperature regime where the efficiency
of the detector increases with rising temperature. 

\end{abstract}

\date{\today}
\maketitle

\section{Introduction}

Detecting the state of a quantum system is an invasive process, which
necessarily modifies the system itself. 
In a continuous measurement description the information on the system's
state is gradually encoded in a classical (macroscopic) signal of
a detector, which at the same time induces a modification of the state
of the system\cite{Clerk2010,Wiseman2010}. In the simplest case of
measuring an observable $A$ of a two-level system, where the detector
distinguishes the two eigenstates of $A$, the process is characterized
by a measurement time, $\tau_{M}$, after which the detector's signals
for the different eigenstates can be resolved from the detector's
noise. From the system's point of view the detector back-action corresponds
to a stochastic component of the state evolution, which asymptotically
drives the system towards one of the measured eigenstates. In average,
this back-action is quantified by the detector-induced decoherence
time, $\tau_{{\rm dech}}$, after which the system is in an incoherent
mixture of eigenstates of $A$. The fundamental disturbance associated
to measurement in quantum mechanics is quantified by the fact that
$\tau_{{\rm dech}}\leqslant\tau_{M}$. When the decoherence rate coincides
with the rate of acquisition of information, back-action is minimal,
which is referred to as quantum-limited detection. This continuous
description of a quantum measurement is in fact appropriate for current
readout methods of a variety of qubits and quantum devices \cite{Morello2010a,Kagami2011,Groen2013,Ward2016}.

The significance of quantum-limited detection is apparent in single
shot measurements, as opposed to averaged measurement results. In
a single shot measurement a quantum-limited detector induces a stochastic
evolution of the system without any decoherence, and therefore a pure
state remains as such during the measurement\cite{Wiseman2010,Korotkov1999,Korotkov}; decoherence appears only as a result of averaging over the detectors's outcome.
This observation is at the basis of a number of techniques for quantum devices control \cite{Wiseman2010,Vandersypen2005,Shapiro2012}, precision measurement \cite{Hosten787,Dixon2009,Dressel2014}, and quantum information processing \cite{Zhang2005,Zilberberg2013,Jordan2015,MeyerzuRheda2014}. The experimental implementation of these techniques besides quantum
optics\cite{Wiseman2010} has been initiated in superconducting qubits
where feedback loops \cite{Vijay2012} and single trajectories mapping
\cite{Weber2014} have been reported. Quantum-limited detection is therefore of interest in solid state
systems at large, where spin, charge, and topologically protected
degrees of freedom are exploited for new quantum devices. A number
of different detection schemes exist in these contexts. For example
charge sensors based on transport through semiconductor devices, like
quantum point contacts (QPCs), are used and proposed as sensors for
e.g. charge \cite{Field1993,Shi2013,Cao2013,Kim2015,Elzerman2003,Petta2004,Oxtoby2006,Ward2016}, spin
\cite{Elzerman2004,Petta2005}, and topologically protected
qubits \cite{aasen2015}.

Motivated by the evolution of measurement process in solid state systems,
we analyze here the effect of interactions on quantum measurement,
focusing on the detector's efficiency. Electron-electron
interactions are generally important in solid state systems. 
Specifically, we consider a charge qubit sensed by a nearby quantum
point contact in the tunneling regime, which directly models charge
sensing in experiments, and can emerge as an effective description of certain detection
schemes of superconducting qubits  \cite{Clerk2006}. We consider two effects of the electrostatic coupling of the
QPC to the charge state of the qubit: (i) a state-dependent tunneling term and
(ii) a state-dependent coupling to the local density \cite{Aleiner1997}. In the absence of interactions, the QPC is a quantum-limited detector for sufficiently low temperature. Both thermal fluctuations and local density couplings drive the detector away from its quantum limit working point \cite{gurvitz,Korotkov1999,Aleiner1997,Averin2005}. 
We find that repulsive electron-electron interactions generically reduce both the rates of induced decoherence and of acquisition of information with respect to their noninteracting counterpart, although in different amounts. This difference is due purely to the local density interaction term, which contributes to decoherence but does not participate in the current and hence provides no information on the system's state. For increasing strong interactions, the renormalization of the rates leads to the need of lower temperatures in order to reach the quantum limit of detection. In this case interactions provide us with a slower
detector. Remarkably, for sufficiently strong interactions we find an intermediate temperature regime
where, as opposed to the noninteracting case, the measurement efficiency
improves with increasing temperature.

The manuscript is organized as follows. In Sec. \ref{sec:Model} we define the model and present the Hamiltonian of the system in the Luttinger formalism. Sections \ref{sec:Decoherence} and \ref{sec:FCS} are devoted to calculating the rates of decoherence and acquisition of information respectively. The decoherence rate is obtained by considering the reduced density matrix of the charge qubit in the presence of the QPC. We show that the two coupling mechanisms to the environment are separable and calculate the tunneling contribution {\it via} a cumulant expansion. The rate of acquisition of information is obtained by considering the full counting statistics of the problem. The effects of electronic interactions on the detection efficiency of the QPC are analyzed in Sec. \ref{sec:QLD}. The conclusions are presented in Sec. \ref{sec:Conclusions}. Lengthy calculations have been relegated to the Appendix.

\section{Model}
\label{sec:Model}

\begin{figure}
\includegraphics[width=0.45\textwidth]{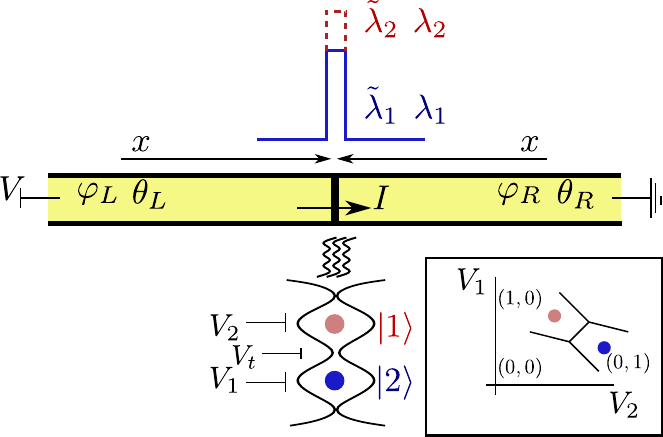}\caption{(Color online) Sketch of the system under consideration. A double quantum dot is
capacitively coupled (wavy lines) to a tunnel junction between two Luttinger liquids.
Two charge configurations $\ket{1}$ (blue solid), $\ket{2}$ (red shady), for an  electron shared
between the two dots  induce different tunnel barriers, hence different currents, through the junction. The electron in the double dot can generically
be in a coherent superposition of these two states, as controlled by external gate voltages, as sketched in the inset.\label{fig:system} }
\end{figure}

We consider a double quantum dot (DQD) which realizes a charge qubit, in proximity
of a QPC. The QPC is formed by a tunneling barrier between two semi-infinite
1D quantum wires consisting of spinless interacting electrons, as
depicted in Fig.\ref{fig:system}. We treat the wires as Luttinger liquids.

The charge configuration of the DQD affects the tunneling of electrons at the QPC between right (R) and left (L) Luttinger liquids. Hence the current through the QPC acts as a charge detector
of the double quantum dot. The total Hamiltonian of our problem consists
of three terms 
\begin{align}
H_{\Sigma}=H_{LL}+H_{QD}+H_{\rm{int}}\,,
\end{align}
where $H_{LL}$ represents the Hamiltonian of both left and right Luttinger
liquids, $H_{QD}$ that of the DQD and $H_{\rm{int}}$ the interaction
between these. If we consider the QPC to be located at $x=0$, 
\begin{align}
H_{LL}=\frac{1}{2\pi}\sum_{j=L,R}v_{g}\int_{-\infty}^{0}\left\{ g\left(\partial_{x}\varphi_{j}\right)^{2}+\frac{1}{g}\left(\partial_{x}\theta_{j}\right)^{2}\right\} {\rm {d}x}\,,\label{eq:LL}
\end{align}
where $\theta_{L\left(R\right)}$ and $\varphi_{L\left(R\right)}$
are the usual charge and phase fields  in the bosonic representation
of the Luttinger liquid on the left (right) side of the QPC, $g$ is the dimensionless interaction parameter
(which for repulsive interactions fulfills $0<g\le1$, being $g=1$ the noninteracting limit) and $v_{g}$
is the group velocity of collective plasmonic excitations. We have chosen the coordinate systems on both left and right such
that $x$ increases from $-\infty$ to zero, where the QPC is located. We have also set $\hbar=1$, which holds hereafter along with $k_B=1$.

The Hamiltonian of the DQD is 
\begin{align}
H_{QD}=\sum_{n=1,2}\varepsilon_{n}c_{n}^{\dagger}c_{n}+\gamma\left(c_{1}^{\dagger}c_{2}+c_{2}^{\dagger}c_{1}\right)\,,\label{eq:dot}
\end{align}
where $c_{n}^{\dagger}\left(c_{n}\right)$ are fermionic operators
of creation-(destruction) of an electron in the n-th quantum dot ($n=1,2$),
$\varepsilon_{n}$ are the electronic level energies (with respect
to the Fermi energy of an external electronic reservoir, which is chosen
to be equal to zero) and $\gamma$ is the tunneling amplitude between
the dot's levels. In the following we assume that the DQD is, besides
the nearby QPC, isolated from the electronic environment, with
a total extra electron shared between the two dots. In this case only
the energy difference $\varepsilon_{2}-\varepsilon_{1}=\varepsilon$
is physical. 

We define further the fields $\theta_{\pm}=1/2\left[\theta_{L}\pm\theta_{R}\right]$,
and $\varphi_{\pm}=1/2\left[\varphi_{L}\pm\varphi_{R}\right]$ and model the interaction term as (cf. Appendix \ref{sec:App_IntH})
\begin{align}
H_{\rm{int}}=\sum_{n=1,2}[a_0\lambda_{n}\partial_{x}\theta_{+}+\tilde{\lambda}_{n}\cos\left(2\varphi_{-}+{\rm {{\rm{eV}}}} t\right)]|_{x=0}c_{n}^{\dagger}c_{n}\,,\label{eq:H_int}
\end{align}
where $\lambda_{n}$ represents the electrostatic coupling between
the quantum dot and the Luttinger liquid leads at $x=0$, and $\tilde{\lambda}_{n}$
characterizes the tunneling at the QPC. Both quantities are assumed to be real and positive, and depend on the state of the DQD, $n$. The parameter
$a_{0}$ is the short-distance cutoff that goes to zero in the continuum
limit. This provides a high-energy cutoff to the model,  $\Lambda_g=v_g/a_0$. Therefore in our further analysis all energies  fulfill $ E\ll \Lambda_g$ and all times $t\gg1/ \Lambda_g$.  ${\rm {{\rm{V}}}}$ is an externally applied voltage bias between
left and right Luttinger liquids. In the limit of weak tunneling which
concerns us here, this potential difference can be described by a
local voltage drop at the QPC site\cite{Egger1998,kane1992transport}.
In what follows we denote the fields evaluated at $x=0$ by simply
omitting the spatial argument.

Note that in the choice of the interaction Hamiltonian we have implicitly
identified the states $\ket{1}\equiv c_{1}^{\dagger}\ket{0}$ and
$\ket{2}\equiv c_{2}^{\dagger}\ket{0}$ as the charge eigenstates
of the measurement device. The detector signal for these two states
and the induced decoherence on their coherent superposition characterize
the tunnel-coupled Luttinger liquids as a detector.

\section{Decoherence}
\label{sec:Decoherence}

In this section we calculate the decoherence rate caused by the tunnel-coupled
Luttinger liquids on the DQD. To do so we assume that at $t=0$ the
DQD is initialized in a coherent state $\ket{\phi_{0}}=\alpha\ket{1}+\beta\ket{2}$
and is decoupled from the detector (i.e. $H_{\rm{int}}=0$). The state of the detector 
is determined by the Hamiltonian $H_{LL}$ in Eq.~\eqref{eq:LL} and by
the temperature $T$ and applied voltage bias ${\rm{V}}$. For $t>0$, the coupling $H_{\rm{int}}$ is suddenly switched on and
the evolution is determined by the DQD interaction with the
QPC. Importantly, we assume a vanishing inter-dot tunneling $\gamma=0$
in Eq.~\eqref{eq:dot} since we are interested in the pure decoherence
induced by the detector (without relaxation processes). Let us note
that physically $\gamma\neq0$ is needed to create the initial coherent superposition
$\ket{\phi_{0}}$, and $\gamma$ can be consistently assumed arbitrarily
small so that the effect of the inter-dot tunneling is negligible
throughout the relevant time scales of system-detector interactions ($t \ll 1/\gamma $). Alternatively, assuming total control of the experimental setup \cite{Cao2013,Kim2015}, $\gamma$ can be set to zero after preparing the coherent state.

To quantify the measurement-induced decoherence we analyze the DQD
reduced density matrix $\rho$, where the degrees of freedom
of the environment (in this case, the LL) have been traced out. The
initial density matrix $\rho(0)=\ket{\phi_{0}}\bra{\phi_{0}}$
at $t=0$ evolves at time $t$ to $\rho_{mn}(t)=e^{-i \left(\varepsilon_{m}-\varepsilon_{n}\right)t}\rho_{mn}(0)\langle U_{n}^{\dagger}(t) U_{m}(t)\rangle$,
where $m,n=1,2$, $H_{QD}\ket{n}=\varepsilon_{n}\ket{n}$, $U_{n}\left(t\right)=\mathcal{T}_{t}\exp\left\{ -i\int_{0}^{t}d\tau{\mathcal{H}}_{{\rm{int}}}^{\left(n\right)}\left(\tau\right)\right\} $,
and $\left\langle ...\right\rangle $ denotes the quantum-statistical average over $H_{LL}$  at temperature $T$. $\mathcal{T}_{t}$ ($\overline{\mathcal{T}}_{t}$) denote time-  (anti-time-) ordering
operators, and ${\mathcal{H}}_{\rm{int}}^{n}(s)=\bra{n}{\mathcal{H}}_{\rm{int}}(s)\ket{n}$, 
where ${\mathcal{H}}_{\rm{int}}$ corresponds to ${H}_{\rm{int}}$ written in the interaction representation
with respect to $H_{LL}$. By using the equation of motion for the
bosonic fields, it can be shown that ${\mathcal{H}}_{\rm{int}}^{n}$ in Eq.~\eqref{eq:H_int}
can be written in terms of phase fields only (see Appendix \ref{sec:App_IntH}),
\begin{equation}\label{eq:Hint}
 {\mathcal{H}}_{\rm{int}}^{n}(t)= -g\frac{\lambda_{n}}{\Lambda_g}\dot{\varphi}_{+}(t)+\tilde{\lambda}_{n}\cos{\left[2\varphi_{-}(t)+{\rm{eV}}t\right]}\,.
\end{equation}

To calculate the time evolution of the reduced density matrix, we first note that
the fields $\varphi_{+}$ and $\varphi_{-}$ commute at equal times,
$\left[\varphi_{+}(t),\varphi_{-}(t)\right]=0,$ which allows in the
following to evaluate their vacuum expectation values separately.
We obtain
\begin{align}
\rho_{mn}(t)=\rho_{nm}^{*}(t)=e^{-i(\epsilon_{m}-\epsilon_{n})t}\rho_{mn}(0)Z_{mn}(t)\tilde{Z}_{mn}(t)\label{eq:rho_mn}
\end{align}
with 
\begin{align}
Z_{mn}&=\left\langle e^{i\frac{g\left(\lambda_{n}-\lambda_{m}\right)\left[\varphi_{+}(t)-\varphi_{+}(0)\right]}{v_{g}}}\right\rangle ,\\
\tilde{Z}_{mn}&= \left\langle \mathcal{U}_n(t,0)^{-1} \mathcal{U}_m(t,0) \right\rangle  ,\label{eq:Ztilde}
\end{align}
with $\mathcal{U}_m(t,0)=\mathcal{T}_{t}e^{-i\tilde{\lambda}_{m}\int_{0}^{t}{\rm d}\tau\cos{\left[2\varphi_{-}(\tau)+{\rm{eV}}\tau\right]}}$.
The two factors $\tilde{Z}_{mn}$ and $Z_{mn}$ correspond to the local density interaction and tunneling induced backaction respectively. 
The only non trivial evolution of the reduced density matrix is in
its off-diagonal terms with $m\neq n$, which take, up to a time-independent prefactor, the form
\begin{align}
Z_{12}&\propto e^{-\left[\Gamma(t)+i\Delta(t)\right]\,t} ,\\
\tilde{Z}_{12}&\propto e^{-[\tilde{\Gamma}(t)+i\tilde{\Delta}(t)]t}\,.\label{eq:Ztilde12}
\end{align}
We identify the respective contributions to the induced energy shift, $\Delta(t)$ and $\tilde{\Delta}(t)$, and decoherence, $\Gamma(t)$ and $\tilde{\Gamma}(t)$. These are generically time dependent quantities. In the following we will focus separately on these contributions to the induced total decoherence $\Gamma_{\rm{tot}}(t)=\Gamma(t)+\tilde{\Gamma}(t)$, which characterize the properties of the QPC as a detector.

\subsection{Local density contribution}

The term $Z_{12}$=$Z_{21}^{*}$ corresponds
to a local change in the electrostatic potential caused by the DQD (see Appendix \ref{sec:App_IntH}),  a fact known to lead to an ``orthogonality catastrophe''  in fermionic systems. The term orthogonality catastrophe refers to the vanishing, in the thermodynamic limit, of the overlap between the system's ground states before and after the change in the potential \cite{AndersonPRL67}.  The average in $Z_{12}$ involves only the $\varphi_+ $-dependent part of the free LL Hamiltonian. Since the latter is quadratic ({\it c.f.} Eqs.~\eqref{eq:Hint} and \eqref{eq:LL}), we can directly write
\begin{align}
Z_{12}(t)= e^{-\frac{1}{2}\left[\frac{g\left(\lambda_{2}-\lambda_{1}\right)}{\Lambda_g}\right]^{2}\left\langle \left(\varphi_{+}(t)-\varphi_{+}(0)\right)^{2}\right\rangle }\,.\label{catastrofe}
\end{align}
The two-point correlation function of  $\varphi_{+}$
is computed in Appendix \ref{integrali}. In the long-time limit $t\gg1/T$, $\Gamma(t)$ is independent of time and we find the local density induced decoherence rate is given by
\begin{align}
\Gamma=\frac{g}{2}\pi T \left[\frac{\left(\lambda_{2}-\lambda_{1}\right)}{\Lambda_g}\right]^{2}\,.\label{eq:decoherenza-catastrofe}
\end{align}
This result is consistent with the known noninteracting ($g=1$) orthogonality exponent in Luttinger systems\cite{Schotte69,Aleiner1997}. Hence we see that for repulsive interactions, the factor  $g<1$ decreases this decoherence rate with respect to the noninteracting case. In a fermionic picture, the orthogonality catastrophe can be seen as a consequence of a ``shake up'' of the Fermi sea due to a change in the local potential. Intuitively, for strong repulsive interactions the electrons will redistribute after the potential change in order to minimize the interaction, consequently minimizing the effect of the shake up.  As expected, larger temperatures lead to a higher decoherence rate. The limit $T \to 0$ leads to the known powerlaw decay of the coherence factor $Z_{mn}$, and hence to logarithmic corrections to the total decoherence rate $\Gamma_{\rm{tot}}(t)$. It should be noted that this result corresponds to an equilibrium ($\rm{{\rm{eV}}}=0$) contribution to the orthogonality catastrophe. This is due 
to the separable character of the reduced density matrix in the weak tunneling limit [{\it c.f.} Eq.~\eqref{eq:rho_mn}]. In this limit nonequilibirum effects are entirely contained in the tunneling term, as calculated in the next subsection.

\subsection{Tunneling term}
\label{subsec:Tunneling-term}

The effect of the change in the transmission of the QPC due to the charge state of the 
DQD is encoded in $\tilde{Z}_{12}=\tilde{Z}^{*}_{21}$. We evaluate this quantity
\textit{via} a cumulant expansion. For simplicity of notation we introduce
the function 
\begin{align}\label{eq:Axi}
A_{\xi(\tau)}(\tau)=\cos{\left[2\varphi_{-}(\tau)+{\rm{eV}}\tau+\frac{\xi(\tau)}{2}\right]}\,,
\end{align}
where $\xi$ is a counting field whose role will be elucidated in
the next section; for the remainder of this section we set $\xi=0$.

We evaluate the time ordered products
to obtain 
\begin{align}
\begin{split}\tilde{Z}_{12}(t) & \approx1+\tilde{\lambda}_{1}\tilde{\lambda}_{2}\int_{0}^{t}{\rm d}\tau\int_{0}^{t}{\rm d}\tau'\langle A_{0}(\tau)A_{0}(\tau')\rangle\\
 & -\tilde{\lambda}_{1}^{2}\int_{0}^{t}{\rm d}\tau\int_{0}^{\tau}{\rm d}\tau'\langle A_{0}(\tau)A_{0}(\tau')\rangle\\
 & -\tilde{\lambda}_{2}^{2}\int_{0}^{t}{\rm d}\tau\int_{\tau}^{t}{\rm d}\tau'\langle A_{0}(\tau)A_{0}(\tau')\rangle\,.
\end{split}
\label{decoerenza}
\end{align}
As shown in  Appendix \ref{App:correlatore}, we can express $\tilde{Z}_{12}(t)$
in terms of the well known time-ordered correlator~\cite{giamarchi2004quantum} $f^{T}(\tau-\tau')\equiv\langle\mathcal{T}_{t}\,e^{2i\varphi_{-}(\tau)}e^{-2i\varphi_{-}(\tau')}\rangle$. In the long time limit $t\gg1/T\,,1/{\rm{eV}}$, we obtain the contribution
to the decoherence
\begin{align}
\tilde{\Gamma}(t)\approx\frac{1}{2}\left(\tilde{\lambda}_{2}-\tilde{\lambda}_{1}\right)^{2}\,\textrm{Re}\left\{ J_{C}\right\} ,\label{eq:decoerenza-tunneling}
\end{align}
 where $J_{C}=\int_{0}^{\infty}{\rm d}s\,f^{T}(s)\cos\left({\rm{eV}}\,s\right)$ [see Eq~\eqref{eq:tildeZmnApp}]. $J_{C}$ is evaluated in Appendix \ref{sec:app-E} [Eq.~\eqref{eq:JC}], and yields the
explicit expression for the (time independent) decoherence rate 
\begin{align}\label{eq:GammaTilde}
\tilde{\Gamma} & =\frac{\left(\tilde{\lambda}_{2}-\tilde{\lambda}_{1}\right)^{2}}{4\Lambda_g}\left(\frac{2\pi T}{\Lambda_g}\right)^{2/g-1}\\
 & \times\frac{\vert\varGamma\left(\frac{1}{g}+i\frac{{\rm{eV}}}{2\pi T}\right)\vert^{2}}{\varGamma\left(\frac{2}{g}\right)}\cosh({\rm{eV}}/2T)\,,\nonumber 
\end{align}
where $\varGamma (x)$ is the gamma function (note the cursive font, not to be confused with the local density induced decoherence rate $\Gamma$).
\begin{figure}
\includegraphics[width=\columnwidth]{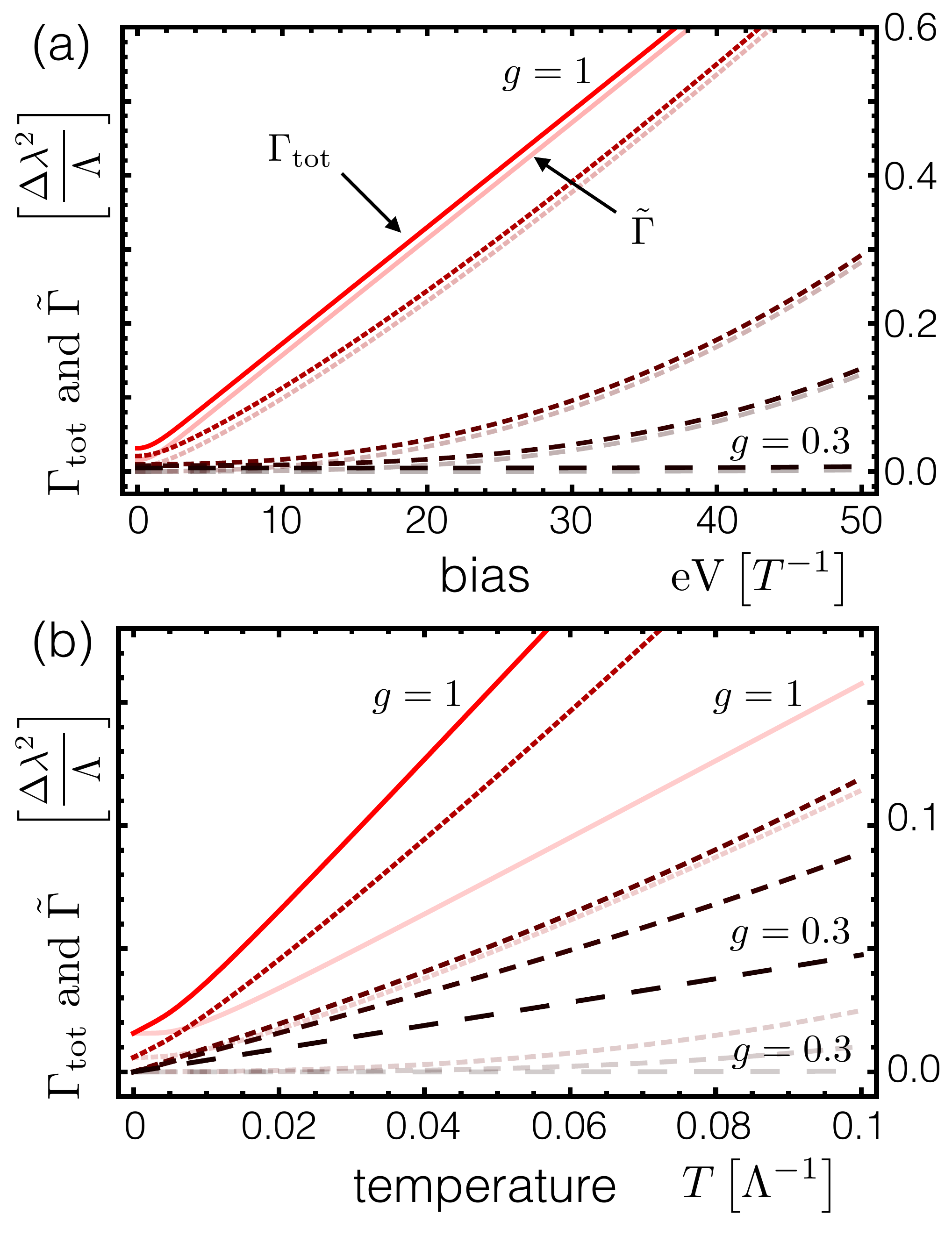}

\caption{(Color Online) Detection induced decoherence: total decoherence $\Gamma_{\text{tot}}=\tilde{\Gamma}+\Gamma$ (solid color) and tunneling induced decoherence $\tilde{\Gamma}$ (shaded color)
as a function of  (a) voltage bias and (b) temperature \textemdash {\it c.f.} Eqs. (\ref{eq:decoerenza-tunneling}),(\ref{eq:decoherenza-catastrofe}). All plots are for
increasing interaction strength $g=1;\,0.9;\,0.6;\,0.5;\,0.3$ from
light to dark red and from continuous to coarsely dashed. The local density induced decoherence rate $\Gamma$ in our model is independent of bias and proportional to temperature, {\it c.f.} Eq.~\eqref{eq:decoherenza-catastrofe}. Hence it produces just a constant shift of the total rate $\Gamma_{\text{tot}}$ in (a) while it modifies the slope in (b). We have
set $\frac{T}{\Lambda_g}=0.01$ in panel (a) and $\frac{{\rm{eV}}}{\Lambda}=0.01$
in panel (b). In all plots $\tilde{\lambda}_{2}-\tilde{\lambda}_{1}=\lambda_{2}-\lambda_{1}=\Delta\lambda$. 
}

\label{fig:fig1} 
\end{figure}
The behavior of  $\tilde{\Gamma}$ is plotted in Fig.~\ref{fig:fig1} as a function of bias voltage and temperature for different values of the interaction strength $g$. As expected, $\tilde{\Gamma}$
increases both as a function of bias and temperature, reflecting the increase in shot and thermal noise respectively. Upon increasing the interaction strength (corresponding to decreasing $g$) however the decoherence generally 
\emph{decreases}, {\it i.e.} electron-electron interactions reduce the
measurement induced backaction. Intuitively, this can be seen as a consequence of an increased ``anti-bunching'' of the electrons with increasing repulsive interactions, which leads to a suppression of tunneling events between the two sides of the QPC.  Since the tunneling processes control the system detector coupling, their suppression results in a reduced back-action onto the DQD. For $g=1$ Eq.~\eqref{eq:GammaTilde} recovers the known result for the decoherence
induced by noninteracting electrons in the tunneling regime. In particular,
for $T\ll {\rm{eV}}$, $\tilde{\Gamma}\approx\frac{{\rm{eV}}}{4\pi v_{F}^{2}}(t_{2}-t_{1})^{2}$\,\,\,
\cite{Korotkov,Averin2005,Aleiner1997}, where $t_n$ are the tunneling strengths introduced in Eq.~\eqref{eq:H_int}. 

\subsection{Total decoherence}
\label{subsec:TotalDec}

\begin{figure}
\includegraphics[width=\columnwidth]{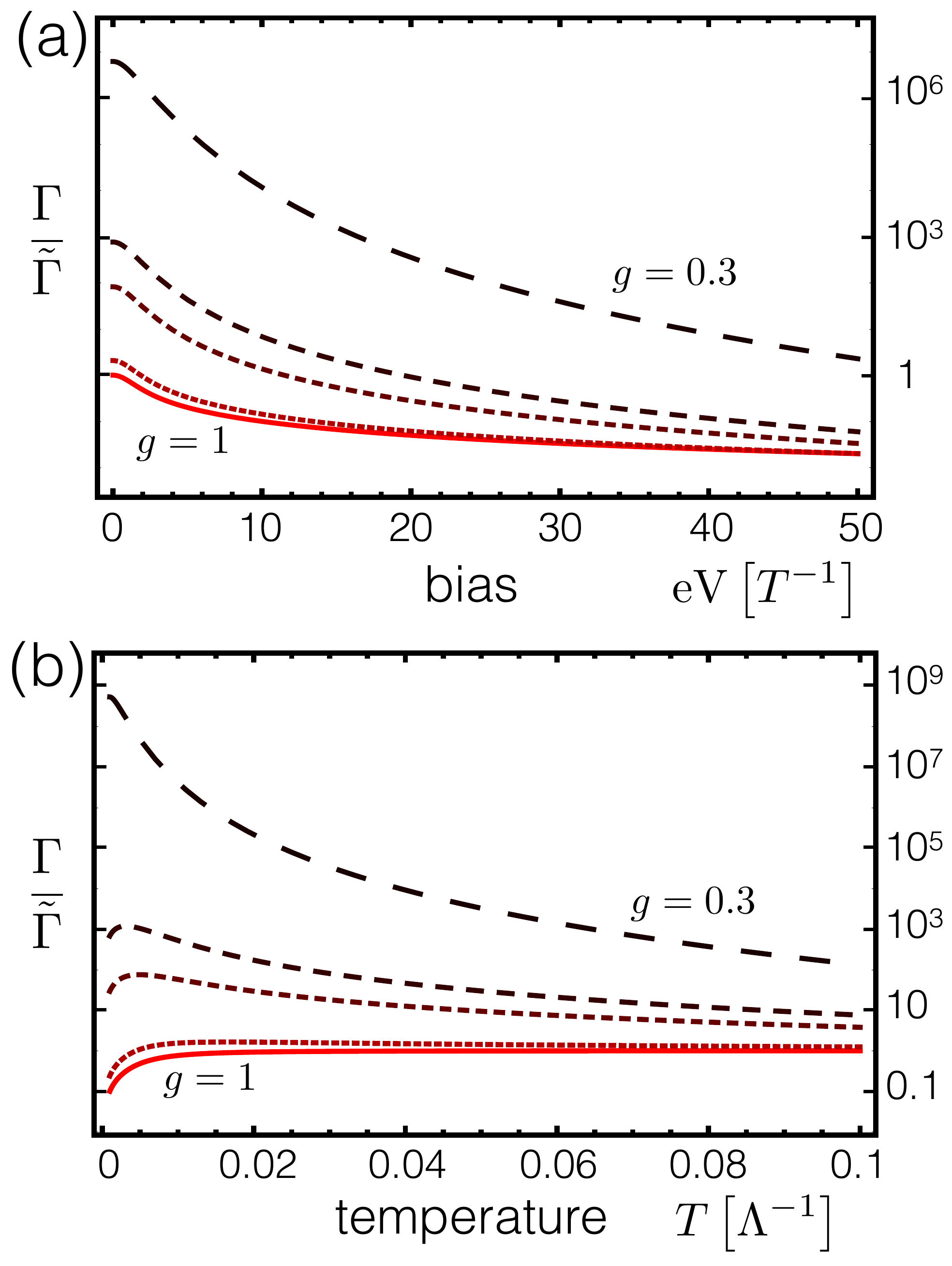}

\caption{(Color Online) Log-plot
of the ratio between local density- and tunneling-induced decoherence
as a function of (a) bias voltage and (b) temperature, {\it cf.} Eqs. (\ref{eq:decoherenza-catastrofe}), (\ref{eq:decoerenza-tunneling}). For
increasing interaction strength $g=1;\,0.9;\,0.6;\,0.5;\,0.3$ from
light to dark red and from continuous to coarsely dashed. $\frac{T}{\Lambda_g}=0.01$ in (a) and $\frac{{\rm{eV}}}{\Lambda_g}=0.01$
in (b), $\tilde{\lambda}_{2}-\tilde{\lambda}_{1}=\lambda_{2}-\lambda_{1}=\Delta\lambda$. 
}
\label{fig:RatesRatio} 
\end{figure}

From the results presented in the two previous subsections, we see that the total decoherence $\Gamma_{\rm{tot}}$ is generically suppressed by increasing repulsive interactions.  This is plotted in Fig.\ \ref{fig:fig1}. Albeit both $\Gamma$ and $\tilde{\Gamma}$
are suppressed, this suppression
is much stronger in the tunneling induced decoherence $\tilde{\Gamma}$,
leading to a variation of the ratio $\Gamma/\tilde{\Gamma}$
by several orders of magnitude depending on interactions as shown
in Fig.\ \ref{fig:RatesRatio}. For instance, for ${\rm{eV}}\ll T\ll\Lambda_g$,
we can analytically approximate 
\begin{align}
\frac{\Gamma}{\tilde{\Gamma}}\simeq\left(\frac{\lambda_{1}-\lambda_{2}}{\tilde{\lambda}_{1}-\tilde{\lambda}_{2}}\right)^{2}\frac{2}{\sqrt{\pi}}\left(\pi\frac{T}{\Lambda_g}\right)^{2-\frac{2}{g}}\sqrt{g}\,,
\end{align}
showing a strong dependence of the ratio on the interactions ($T\ll\Lambda_g$). The strength of this effect is suppressed at larger
voltage bias or temperature.

The reason behind this behavior is the decreasing strength
of the tunneling term in Eq.~\eqref{eq:H_int} as compared to the
local density interaction one in the effective low energy behavior.
The decoherence is generically dictated by the low frequency correlations
of the bath coupled to the system \cite{Leggett1987,Weiss2012} (in our case the
Luttinger liquid detector), hence by the dynamics of the low frequency modes of the bath. 
When tracing out the fast (high energy) modes of the Luttinger liquid detector, the effective low energy tunneling term is suppressed as compared to the local density term. This suppression is more prominent for stronger repulsively interacting systems, as shown by Kane and Fisher \cite{Fisher1997,kane1992transport,KanePRB92}. This leads to a divergent ratio $\Gamma/\tilde{\Gamma}\rightarrow\infty$ when
$T\rightarrow0$ and ${\rm{eV}}\rightarrow0$ simultaneously. When going to higher temperatures
or higher voltages the relative strength of the two contributions
evolves towards comparable values (set by the bare constants $\tilde{\lambda}_{n}$,
$\lambda_{n}$).

\section{Full counting statistics and rate of acquisition of information}
\label{sec:FCS}

The backaction of the detector on the measured system has to be compared
with the ability of the detector to discriminate the different charge
states of the DQD. For a given charge eigenstate ($n=1,2$)
of the double dot, the response of the detector is fully characterized
by the probability distribution $\mathcal{P}_{n}(N,t)$ of a charge
$q=eN$ to be transmitted through the tunnel junction in a fixed time
interval $t$. The rate of acquisition of information on the charge
state of the DQD is quantified by the statistical quantity \cite{Averin2005}
\begin{align}
\mathcal{M}(t)\equiv e^{-W(t)t}\equiv\sum_{N}\sqrt{\mathcal{P}_{1}(N,t)\mathcal{P}_{2}(N,t)},
\end{align}
which measures how distinguishable the two distributions are.

The probability distribution $\mathcal{P}_{n}(N,t)$ is equivalently
and conveniently characterized by the corresponding generating function $\chi_{n}(\xi,t)\equiv\sum_{N}\mathcal{P}_{n}(N,t)e^{i\xi N}$, the so called Full Counting Statistics (FCS). The generating function  can be expressed directly in terms of quantum averages of
the tunneling operator \cite{Levitov2004}
\begin{align}\label{eq:GenFunc}
\chi_n(\xi,t)=e^{W_n(\xi,t)\,t}=\left\langle{\mathcal{T}}_K\exp\left\{i\tilde{\lambda}_n\int_{{\mathcal{C}}_K} d \tau A_{\xi(\tau)}(\tau)\right\}\right\rangle\,,
\end{align}
where the time ordering $\mathcal{T}_{K}$ occurs on the Keldysh contour
$\mathcal{C}_{K}$ and $\xi(\tau)=\pm\xi$ is the counting field introduced
in Eq.~\eqref{eq:Axi}. The FCS of interacting electrons is known in some cases, e.g. in quantum dots or diffusive conductors \cite{Bagrets2006}. Here wre consider instead Luttinger liquids. In the present situation of a tunneling Hamiltonian, the counting field enters as a phase of the tunneling operators~\cite{Levitov2004}
$t\rightarrow te^{i\xi(t)/2}$, $t^{*}\rightarrow t^{*}e^{-i\xi(t)/2}$. In this case the counting field is a pure quantum field, {\it i.e}. $\xi(\tau)=\pm\xi$ is anti-symmetric
on the forward and backward branch of the Keldysh contour. 

The generating
function $\chi_{n}(\xi)$ in Eq.\ \eqref{eq:GenFunc} is a generalization of $\tilde{Z}_{nn}(t)$ in  Eq.~\eqref{eq:Ztilde} which includes the quantum field $\xi(\tau)$. Similarly as we did in the previous section for $\tilde{Z}_{12}(t)$, we evaluate $\chi_{n}(\xi)$ to second order in a cumulant expansion. In the long-time limit ($t\gg1/T\,,1/{\rm{eV}}$), the Markovian nature of the electron transfer
processes guarantees that the leading contribution to the cumulant
generating function is linear in time, {\it i.e.} $W_{n}(\xi,t)\approx W_{n}(\xi)$
is independent of $t$. We obtain
\begin{align}
W_{n}(\xi)=\tilde{\lambda}_{n}^{2}\left[\left(\cos\xi-1\right){\rm Re}\left\{ J_{C}\right\} -i\sin\xi{\rm Im}\left\{ J_{S}\right\} \right],\label{eq:FCS}
\end{align}
where  $J_{S}=\int_{0}^{\infty}{\rm d}s\,f^{T}(s)\sin\left({\rm{eV}}\,s\right)$ is calculated in Appendix \ref{sec:app-E}. In this limit the rate of acquisition of information
$W(t)$ can be expressed directly in terms of $W_{n}(\xi)$ as \cite{Averin2005}
\begin{align}
W(t)=W\approx-\frac{1}{2}\min_{x\in\mathbb{R}}\left\{ W_{1}(-ix)+W_{2}(+ix)\right\} ,
\end{align}
where $W$ can be directly evaluated from Eq.~\eqref{eq:FCS} to be
$W\approx{\rm Re}\left\{ J_{C}\right\} \tilde{\lambda}_{+}-\sqrt{\left({\rm Re}\left\{ J_{C}\right\} \tilde{\lambda}_{+}\right)^{2}-\left({\rm Im}\left\{ J_{S}\right\} \tilde{\lambda}_{-}\right)^{2}}$,
with $\tilde{\lambda}_{-}\equiv(\tilde{\lambda}_{2}^{2}-\tilde{\lambda}_{1}^{2})/2$,
$\tilde{\lambda}_{+}\equiv(\tilde{\lambda}_{2}^{2}+\tilde{\lambda}_{1}^{2})/2$
and $0<\lambda_{-}<\lambda_{+}$. From the expression for $J_{S}$ in Appendix \ref{sec:app-E},
we obtain 
\begin{align}
W={\rm Re}\left\{ J_{C}\right\} \left[\lambda_{+}-\sqrt{\lambda_{+}^{2}-\left(\tanh[{\rm{eV}}/2T]\,\lambda_{-}\right)^{2}}\right].\label{eq:information}
\end{align}
In the next section we discuss the implications of this result for
the quantum measurement process. We note here that the acquisition of
information is independent of the local density interaction contributions
(parametrized by $\lambda_{n}$), since these do not affect the current
and hence do not contribute to the gain of knowledge about the charge
state of the DQD.

\section{Effects on quantum-limited detection}
\label{sec:QLD}

The efficiency of the quantum measurement is characterized by the
ratio 
\begin{align}
Q\equiv W/\Gamma_{{\rm tot}}=W/(\Gamma+\tilde{\Gamma})\leqslant1\,.
\end{align}
This definition takes only into account the decoherence on the measured system due to the measurement process, following the approach used for non-interacting detectors. $Q=1$ corresponds to a quantum-limited detector. External, system-dependent decoherence mechanisms are outside the scope of this paper.

The efficiency $Q$ is properly defined for sufficiently long times $t>1/T\,,1/{\rm{eV}}$, where $Z_{mn}(t)$ and $\tilde{Z}_{mn}(t)$ are exponentially decaying in time and $W(t)=W$. With the help of Eqs.~\eqref{eq:decoerenza-tunneling} 
and \eqref{eq:information} we conveniently rewrite $Q$
as 
\begin{align}
Q=\frac{\frac{1+\eta^{2}}{2\eta^{2}}\left[1-\sqrt{1-\left(\frac{2\eta}{1+\eta^{2}}\tanh({\rm{eV}}/2T)\right)^{2}}\right]}{\left(1+\Gamma/\tilde{\Gamma}\right)}\,,\label{eq:Q}
\end{align}
where $\eta=(\tilde{\lambda}_{2}-\tilde{\lambda}_{1})/(\tilde{\lambda}_{1}+\tilde{\lambda}_{2})$
characterizes how strong the electron tunneling is influenced by the
different occupation of the DQD. It can be shown that $Q\le1$ and finite for $\eta \rightarrow 0$. 

\begin{figure}[t]
\includegraphics[width=\columnwidth]{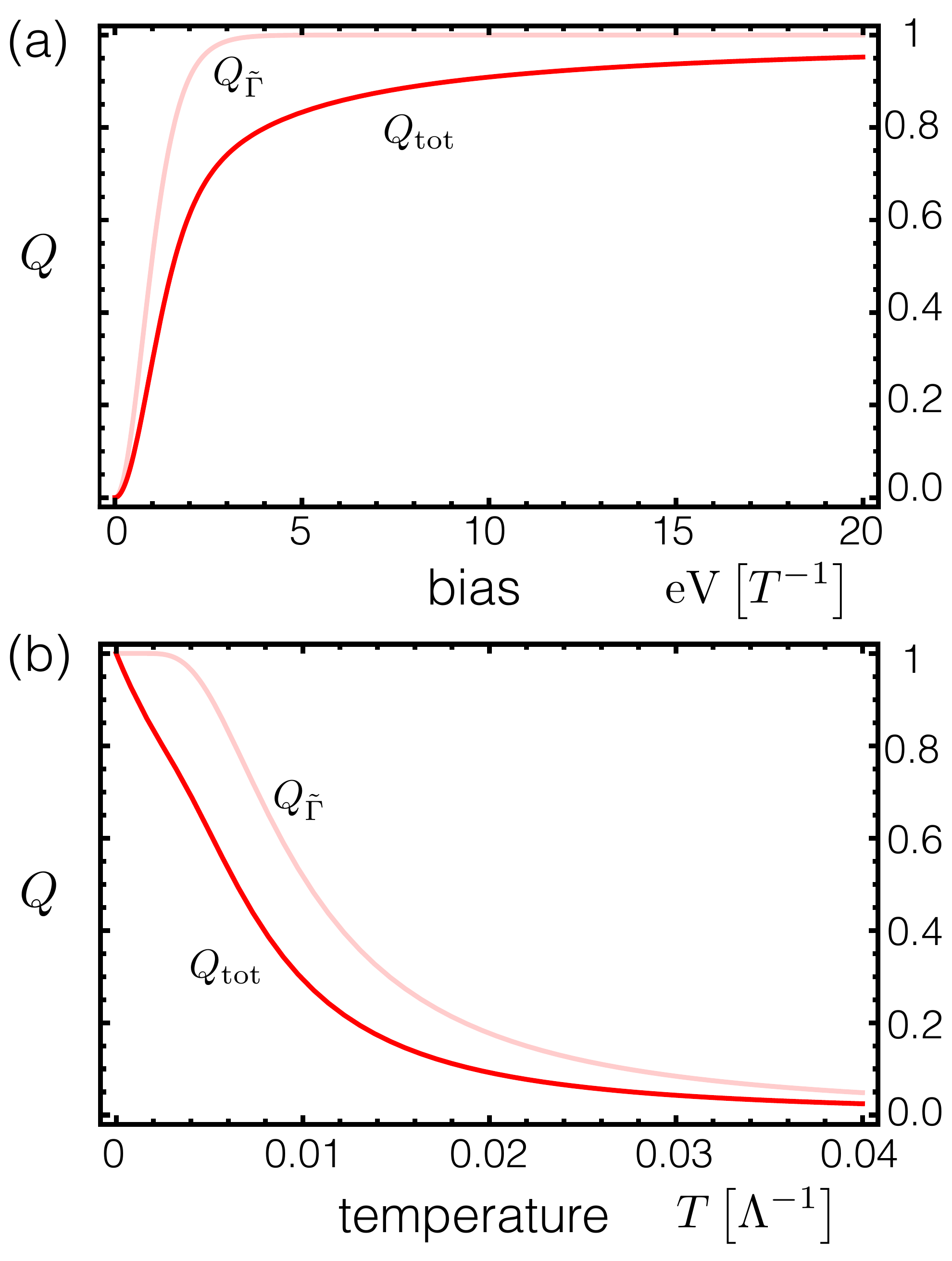}

\caption{(Color Online) Detector efficiency for the noninteracting case. Total efficiency $Q_{\rm{tot}}=W/\Gamma_{\rm{tot}}$ (solid line) and efficiency without decoherence due to the local density interaction, $Q_{\tilde{\Gamma}}=W/\tilde{\Gamma}$ (shaded line) as function of (a) bias voltage and (b) temperature. $Q_{\tilde{\Gamma}}$ is independent of $g$ and therefore valid also for the interacting case. With $\eta=0.5$, $\tilde{\lambda}_{2}-\tilde{\lambda}_{1}=\lambda_{2}-\lambda_{1}$,  $\frac{T}{\Lambda_g}=0.01$ in (a), and $\frac{{\rm{eV}}}{\Lambda_g}=0.01$ in (b).\label{fig:fig4} }
\end{figure}

From Eqs.~\eqref{eq:information} and  \eqref{eq:decoerenza-tunneling}, we note that $W\propto\tilde{\Gamma}\propto {\rm Re}\left\{ J_{C}\right\}$, where all interaction effects (characterized by $g$) are contained in the function ${\rm Re}\left\{ J_{C}\right\}$. Therefore, in the absence of  a local density contribution ($\lambda_1=\lambda_2$, so $\Gamma=0$) $Q$
is independent of $g$ and hence interactions have no effect on the quality
of the detection process. The efficiency $Q$ for $\Gamma=0$ is plotted in Fig. \ref{fig:fig4}(a) and (b). In particular, for $T\ll {\rm{eV}}$ (and $T \neq 0$), the detector is quantum
limited, $Q\rightarrow1$, and remains such in the presence of interactions.  Repulsive
interactions do have an effect in reducing the backaction (cf. Fig.
\ref{fig:fig1}), but the rate of acquisition of information is reduced
by an equal amount. All in all in absence of a local density interaction, 
interactions leave the detector still quantum-limited, but slow down
the detection process. As for non-interacting QPC detectors, the efficiency
of the detection is controlled only by ${\rm{eV}}/T$ , and in the limit
of high temperature, thermal fluctuations induce unwanted backaction
unaccompanied by information gain, driving the detector away from
its quantum limit {[}cf. Fig. \ref{fig:fig4}(a) and \ref{fig:fig4}(b){]}.

\begin{figure}[t]
\includegraphics[width=\columnwidth]{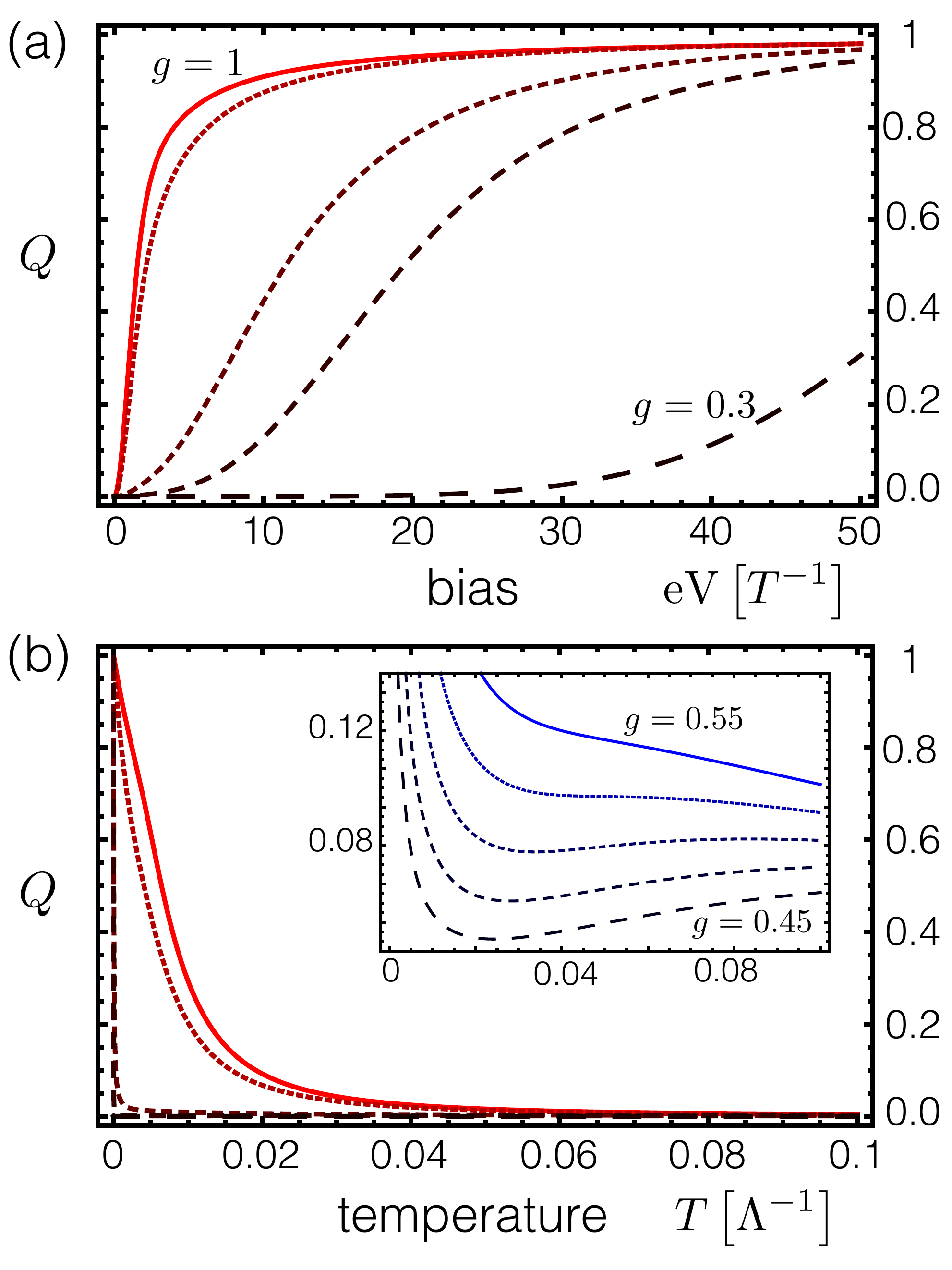}

\caption{(Color Online) Total detector efficiency $Q$ for the interacting case, as function of the applied bias (a) and temperature (b). Different curves from light to dark
red and from continuous to coarsely dashed are for increasing interaction
strengths, $g=1;\,0.9;\,0.6;\,0.5;\,0.3$. The figures show strong dependence of
$Q$ on $g$ once orthogonality effects are considered. The inset in (b) shows a zoom in of the regime of crossover between monotonous and non-monotonous
temperature dependence for $g\sim0.5$ ($g=0.55;\,0.525;\,0.5;\,0.475;\,0.45$
from light to dark blue and from continuous to coarsely dashed). We
used $\eta=0.5$, $\tilde{\lambda}_{2}-\tilde{\lambda}_{1}=\lambda_{2}-\lambda_{1}$, and  $\frac{T}{\Lambda_g}=0.01$ in (a), and  $\frac{{\rm{eV}}}{\Lambda_g}=0.01$
in (b).\label{fig:fig5} }
\end{figure}

Therefore the local density interaction is essential to appreciate the effect
of interactions. Once it is taken into account, lower temperatures are required to bring the detector to the quantum limit, even in the absence of
interactions (see Fig.~\ref{fig:fig4}). This is due to the fact that this term provides no information gain, but still induces
decoherence onto the system~\cite{Aleiner1997}. We showed in Sec.
\ref{subsec:Tunneling-term} that both decoherence due to the local density interactions \emph{and }due to tunneling are diminished by interactions,
but in a very unequal way. The suppression of the tunneling-induced decoherence $\tilde{\Gamma}$ due to electron-electron interactions is much more pronounced than that of the decoherence caused by the local density contribution, $\Gamma$. Since $W\propto\tilde{\Gamma}$, the acquisition of information is suppressed in the same manner. This leads to a strong suppression of the measurement efficiency
$Q$ [Eq.\ \eqref{eq:Q}] for repulsive interactions with respect to the noninteracting case. Interaction effects
do not eliminate the monotonously increasing dependence on the voltage
bias of $Q$ {[}cf. Fig.\ \ref{fig:fig5} (a){]}, but can delay the saturation
to the quantum limit $Q=1$ to very high voltages or very low temperatures
for strongly repulsively interacting systems. 

Surprisingly the temperature dependence of $Q$ shows an interesting
nonmonotonous feature depending on interactions. For a noninteracting
system, $Q$ is a monotonously decreasing function of temperature, reflecting the
fact that increasing thermal fluctuations induce extra decoherence
without a corresponding gain of information about the system's state
{[}cf. Fig. \ref{fig:fig4} (b){]}. However we find that for strong interactions and at high temperatures with respect to the bias, $Q$ increases with $T$ in an intermediate regime. Specifically,
for $0<{\rm {eV}}\ll T<\frac{\Lambda_{g}}{2\pi}\left(\frac{\varGamma\left(\frac{1}{g}\right)^{2}\left(\tilde{\lambda}_{1}-\tilde{\lambda}_{2}\right)^{2}}{g\varGamma\left(\frac{2}{g}\right)\left(\lambda_{1}-\lambda_{2}\right)^{2}}\right)^{\frac{g}{2g-2}}\ll\Lambda_{g}$,
we have 
\begin{align}
Q\simeq\frac{(\tilde \lambda_1 - \tilde \lambda _2 )^2\, \varGamma\left(\frac{1}{g}\right)^{2}}{( \lambda_1 -  \lambda _2 )^2 \left(\eta^{2}+1\right)g\varGamma\left(\frac{2}{g}\right)}\left(\frac{2\pi}{\Lambda_g}\right)^{2/g-2}\left[{\rm{eV}}\right]^{2}T^{\frac{2}{g}-4}\,.
\end{align}
The expression shows a crossover between an increasing and a decreasing
function of $T$ for $g\simeq1/2$ {[}cf. also inset in Fig. \ref{fig:fig5}
(b){]}. This feature emerges from the competition between two effects
of increasing temperature: (i) an increase of thermal fluctuations
and (ii) a increasing prominence of the tunneling term compared to
the local density one {[}cf. Fig. \ref{fig:RatesRatio} (b){]}. To highlight these competing effects we can write the efficiency in Eq. (\ref{eq:Q}) as $Q=Q_0/(1+\Gamma/\tilde{\Gamma})$, so that $Q_0$ is a monotonously decreasing function of temperature. 
At low energies, decoherence is dominated by the local density term due to supression of tunneling, and we can roughly write $Q\sim Q_0 (\tilde{\Gamma}/ \Gamma)$. While
the thermal fluctuations reduce $Q$, the growing prominence of the
tunneling term increases the weight of the ``information carrying'' part of the interaction
Hamiltonian, hence increasing $Q$. When (ii) is dominant compared
to (i), $Q$ increases with temperature. This is controlled by the
parameters of the detector. In particular, since the temperature dependence
of the relative strength between local density and tunneling
contributions in the detector is strong for strong interaction, the
increasing behavior of $Q$ with $T$ is possible only for sufficiently
small $g$. The inset in Fig. \ref{fig:fig5} (b) shows a zoom into
the critical regime of the crossover between monotonous and non-monotonous
temperature dependence.

\section{Conclusions and Outlook}
\label{sec:Conclusions}

In this paper we analyzed the effects of interactions on the efficiency
of quantum detection. We executed our analysis for two voltage biased
electron reservoirs connected by a tunnel junction, whose current
serves as a charge detector of a proximate charge qubit. We included
electron-electron interactions by modeling the leads as Luttinger liquids
and incorporated the effects of local density fluctuations due to the charge qubit, besides its effect on the tunneling amplitude. The model is of interest both
for charge sensing schemes used in experiments and as a theoretical
paradigm case study. 

We found that interactions reduce the induced decoherence
on the measured system, along with the rate of acquisition of information. In the absence of a local density interaction term, both acquisition of information and tunneling induced decoherence are suppressed in the same manner by interactions. In this case interactions do not alter the efficiency of the detector, which tends to be quantum-limited at low temperature, but slow down its response. Once the local density induced decoherence is considered, interactions do play a role for the efficiency, reducing it with respect to the non-interacting case. 

The relative contributions of tunneling and local density induced decoherence are strongly affected by interactions, and
the local density contribution can dominate at low temperature and voltage bias for strong interactions. This is a consequence of the downwards renormalization of the tunneling term for repulsive interactions at low energies. The same renormalization is responsible for the slower rate of acquisition of information in the interacting case. This renormalization is less pronounced for increasing energy, resulting in a tendence to an increased acquisition of information rate.  As a result of the interplay between these effects, we have identified an intermediate temperature regime where, for sufficiently strong interactions
($g\lesssim1/2$), the detector efficiency \textit{increases} with temperature.
This has to be contrasted with the weakly interacting case where increasing
thermal fluctuations monotonously reduce the detector's efficiency.
As a function of the voltage bias, repulsive interactions delay the quantum limit
$Q=1$ to increasingly higher voltages (or lower temperatures). This is a pure consequence of the local density interaction.

Our models captures the effects of interactions in the simplest experimentally relevant configuration. As such, it has limitations and poses interesting future challenges, which we outline briefly here. Our results allow us to assess the efficiency of the detector due to processes inherent to the measurement itself, which are unavoidable as long as the system is coupled to the detector for readout. The readout efficiency will also be affected by other external decoherence mechanisms extraneous to the measurement process. These have to be dealt with separately and are system-specific. For instance, one can come up with more efficient qubit designs or environment engineering to minimize the coupling to specific decoherence sources. Moreover,  in our model we assumed full control of the tunneling matrix element between the dots, which allowed us to set $\gamma$ smaller than all the other energies in the model after preparing the initial coherent state. Our results are valid for $t\ll1/\gamma$ such that we can 
effectively consider $\gamma=0$. Experimentally, the required degree of control is available for charge qubits, though with more sophisticated designs than a double quantum dot \cite{Shi2013,Cao2013,Kim2015}, and for spin qubits whose spin state is read by quantum point contacts via spin-to-charge conversion mechanisms \cite{Petta2004,Oxtoby2006}. The protocol we analyze has to be considered as a test for the detector's properties. In fact, based on results for noninteracting systems~\cite{Korotkov}, there are reasons to expect that the parameters for which the detector is found to be quantum-limited in our manuscript, make the detector quantum-limited also in presence of inter-dot tunneling. The argument is that the efficiency is a property of the measurement process and the detector, not of the qubit's dynamics. A proper analysis  of the dot-detector coupling in the presence of finite inter-dot tunneling is a key future point to address, especially because this is the regime where measurement-based control of the 
qubit dynamics can operate. One can anticipate for instance that pure decoherence will be accompanied by relaxation processes. We have also modeled the DQD as single level dots with single occupation, which is the simplest experimentally relevant case. Considering double occupation requires a treatment with a larger qubit Hilbert space and therefore addressing the coherences of different off-diagonal terms, which is outside of the scope of this manuscript but would be an interesting follow up problem. Lastly, the nonmonotonic behavior of the efficiency $Q$ with temperature is present for strong interactions, $g\lesssim1/2$. Although this is an experimentally challenging regime, recent experiments in different platforms have shown evidence of Luttinger liquid behavior with interactions up to $g\approx 0.2$~\cite{Ishii2003,Levy2012,Li2015}.

\acknowledgments

The work is supported
by Deutsche Forschungsgemeinschaft under the Grant No. RO4710/1-1.
and the SFB 658 (A.B.). S.V. K. acknowledges support from ERC-StG OPTOMECH.

 \bibliographystyle{nourl_apsrev}

\begin{widetext}

\appendix

\section{Coupling Hamiltonian}

\label{sec:App_IntH} We model the electrostatic coupling of the DQD to
the interacting QPC to include two effects: a
coupling of the electron on the DQD to the local electronic density
at the end ($x=0$) of the two Luttinger liquid leads that depends on the charge state of the DQD, and a state-dependent tunneling between the
two sides of the QPC. To derive the interaction Hamiltonian Eq.\ \eqref{eq:H_int}
we start from the fermionic representation 
\begin{align}
\begin{split}H_{\rm{int}} & =\sum_{n,j,c}\alpha_{n}:\Psi_{c,j}^{\dagger}\Psi_{c,j}(0,t):c_{n}^{\dagger}c_{n}\\
 & +\left[t_{n}:\Psi_{2,L}^{\dagger}\Psi_{1,R}(0,t):+t_{n}^{\ast}:\Psi_{2,R}^{\dagger}\Psi_{1,L}(0,t):\right]c_{n}^{\dagger}c_{n}\,,
\end{split}
\label{eq:H_int_APP}
\end{align}
where $\Psi_{c,j}^{\dagger}\left(x,t\right)$ {[}$\Psi_{c,j}\left(x,t\right)${]}
creates (anihilates) an electron at position $x$ and time $t$, with
chirality $c=1,2$ and on side $j=L,R$ {[}note that $c=1$ ($c=2$)
indicates moving towards (away from) the QPC{]}. $n=1,2$ indicates
the state of the DQD and $:\dots:$ normal ordering. These fermionic
fields can be written in terms of the bosonic operators as 
\begin{align}
\Psi_{c,j}=\frac{\eta_{c,j}}{\sqrt{2\pi a_{0}}}e^{ick_{F}x}e^{\mp i{\rm {{\rm{eV}}}t/2}}e^{i\left(c\theta_{j}+\varphi_{j}\right)}\label{eq:Psis_APP}
\end{align}
where all fields are evaluated at $\left(x,t\right)$. $\eta_{c,j}$
are Klein factors, $k_{F}$ the Fermi momentum, the $\mp$ in the
exponential corresponds to $R$ (-) and $L$ (+) and ${\rm {{\rm{eV}}}=\mu_{L}-\mu_{R}}$.

We consider the tunneling term as a perturbation on the two (L,R)
disconnected LL systems. Without tunneling, the QPC acts as a strong
impurity which imposes that the density fluctuations vanish at $x=0$.
This boundary condition results in \cite{giamarchi2004quantum,KanePRB92}
\begin{align}
\theta_{L}\left(x=0,t\right)=\theta_{R}\left(x=0,t\right)=0\,.
\end{align}
Using this condition together with $t_{n}=t_{n}^{*}$ and substituting
Eq.~\eqref{eq:Psis_APP} into Eq.~\eqref{eq:H_int_APP} we obtain straightforwardly
the second term in Eq.~\eqref{eq:H_int}, with $\tilde{\lambda}_{n}=t_{n}/(\pi a_{0})$.

It is furthermore convenient to write the bosonic fields in the interaction
representation, in which the bosonic fields evolve according to the
free Hamiltonian $H_{{\rm {LL}}}$ Eq.\ \eqref{eq:LL} and switch to the description in terms of sum and difference fields $\theta_{\pm}=1/2\left[\theta_{L}\pm\theta_{R}\right]$,
and $\varphi_{\pm}=1/2\left[\varphi_{L}\pm\varphi_{R}\right]$. Using the commutators ($\alpha=\pm$)
\begin{align}
\begin{split}\left[\theta_{\alpha}(x),\varphi_{\alpha'}(x')\right] & =\frac{i\pi}{4}\sgn{(x-x')}\delta_{\alpha,\alpha'}\,,\\
\left[\theta_{\alpha}(x),\partial_{x'}\varphi_{\alpha'}(x')\right] & = - \frac{i\pi}{2}\delta(x-x')\delta_{\alpha,\alpha'}\,,
\end{split}
\end{align}
we obtain the free Heisenberg equation of motion 
\begin{align}
\dot{\varphi}_{\pm}(x,t) & = -\frac{v_{g}}{g}\partial_{x}\theta_{\pm}(x,t)\,,\label{eq:phidot}
\end{align}
\begin{align}
\dot{\theta}_{\pm}(x,t)=- gv_{g}\partial_{x}\varphi_{\pm}(x,t)\,.
\end{align}

The first term in Eq.~\eqref{eq:H_int_APP}, which is the density-density
electrostatic interaction between the dot and the LL at $x=0$, can
easily be bosonized using the identity $\rho_{j}=\sum_{c}:\Psi_{c,j}^{\dagger}\Psi_{c,j}:=\partial_{x}\theta_{j}/\pi$
for the normal ordered density (\textit{i.e.} the density of charge
fluctuations). Using Eq.~\eqref{eq:phidot} we can express $\partial_{x}\theta_{j}$
in terms of $\dot{\varphi}_{j}$ and we obtain the first interaction
term in Eq.~\eqref{eq:H_int} with $\lambda_{n}=\frac{2\alpha_{n}}{\pi a_0}$.

Finally, Eq. \eqref{eq:phidot} allows
us to write $\tilde{H}_{\rm{int}}^{n}=\bra{n}\tilde{H}_{\rm{int}}\ket{n}$ in
terms of phase fields only 
\begin{align}
\tilde{H}_{\rm{int}}^{n}(t)= - g \frac{\lambda_{n}}{\Lambda_g}\dot{\varphi}_{+}(t)+\tilde{\lambda}_{n}\cos{\left[2\varphi_{-}(t)+{\rm{eV}}t\right]}\,,
\end{align}
with $\Lambda_g=v_g/a_0$ the high energy cutoff.

\section{Calculation of  \texorpdfstring{$Z(t)$}{Z(t)}}

\label{integrali}

The detector's contribution to the evolution of the off diagonal terms
of the density matrix is expressed in terms of averages of the detector's
fileds in Eqs. (\ref{catastrofe},\ref{decoerenza}).  We compute here these averages $\left\langle \left(\varphi_{\pm}(t)-\varphi_{\pm}(0)\right)^{2}\right\rangle $. An alternative calculation of the same average can be found in Ref.  \onlinecite{giamarchi2004quantum}. In order to proceed,
it is useful to write the phase fields in terms of bosonic operators
in the interacting basis, 
\begin{align}
\varphi_{\pm}\left(x,t\right)=i\sqrt{\frac{\pi}{2Lg}}\sum_{k\neq0}\frac{e^{-\left|k\right|a_{0}/2}}{\sqrt{\left|k\right|}}\left[e^{-ipx}b_{k,\pm}^{\dagger}\left(t\right)-e^{ipx}b_{k,\pm}\left(t\right)\right]\,,\label{eq:phi_Fourier}
\end{align}
where we have introduced the standard high-energy cut-off $\exp\left(-\left|k\right|a_{0}/2\right)$.
The bosonic creation {[}destruction{]} operators $b_{k}^{\dagger}\left(t\right)$
{[}$b_{k}\left(t\right)${]} are of the form 
\begin{eqnarray}
 & b_{k,\pm}^{\dagger}\left(t\right)=b_{k,\pm}^{\dagger}\left(0\right)e^{iv_{g}t\left|k\right|}\,,\nonumber \\
 & b_{k,\pm}\left(t\right)=b_{k,\pm}\left(0\right)e^{-iv_{g}t\left|k\right|}\,,
\end{eqnarray}
where $b_{k,\pm}^{\dagger}\left(0\right)$ and $b_{k,\pm}\left(0\right)$
fulfill standard bosonic commutation relations ($\alpha,\alpha'=\pm$)
\begin{align}
\left[b_{k,\alpha}(0),b_{k',\alpha'}^{\dagger}(0)\right]=\delta_{k,k'}\delta_{\alpha,\alpha'}\,.
\end{align}
The free Hamiltonian of the Luttinger liquid in this representation
is simply 
\begin{align}
H_{LL}=\sum_{\alpha=\pm;k=0}^{\infty}v_{g}\left|k\right|b_{k,\alpha}^{+}\left(0\right)b_{k,\alpha}\left(0\right)\,,
\end{align}
and the vacuum expectation value 
\begin{align}
\langle b_{k,\alpha}^{+}\left(0\right)b_{k',\alpha'}\left(0\right)\rangle=n_{b}\left(k\right)\delta_{k,k'}\delta_{\alpha,\alpha'}\,,\label{eq:nb}
\end{align}
where $n_{b}\left(k\right)=\left[\exp\left(v_{g}\left|k\right|/T\right)-1\right]^{-1}$
is the usual Bose-Einstein distribution at temperature $T$, with
$k_{B}=1$.

Using Eqs.~\eqref{eq:phi_Fourier}-~\eqref{eq:nb}, we can perform
the vacuum expectation value by going to the continuum limit
\begin{align}
\begin{split}\left\langle \left[\varphi_{\pm}(\tau)-\varphi_{\pm}(\tau')\right]^{2}\right\rangle  & =\frac{1}{g}I(\tau-\tau')\\
I(s)= & \mathcal{P}\int_{0}^{\infty}\frac{{\rm d}k}{k}e^{-a_{0}k}\left[2n_{b}(k)+1\right]\left[1-\cos\left(v_{g}ks\right)\right]\,.\label{eq:THE_{i}ntegral}
\end{split}
\end{align}
where $\mathcal{P}$ denotes the principal value of the integral.

We can divide the integral into a zero temperature quantum term, and
a thermal term proportional to $n_{B}(v_{g}k)$. The quantum term
can be calculated to be
\begin{align}
\mathcal{P}\int_{0}^{\infty}\frac{dk}{k}e^{-a_{0}k}\left[1-\cos\left(v_{g}k\tau\right)\right]=\frac{1}{2}\log\left[1+\left(\frac{v_{g}\,\tau}{a_{0}}\right)^{2}\right]\,.\label{eq:I_quantum}
\end{align}
For the thermal contribution in turn we obtain 
\begin{align}
\mathcal{P} & \int_{0}^{\infty}\frac{dk}{k}e^{-a_{0}k}\left[1-\cos\left(v_{g}k\,\tau\right)\right]n_{B}(v_{g}k)\label{eq:I_thermal}\\
= & \frac{1}{2}\log\left(\frac{\varGamma(1+\frac{a_{0}}{\beta v_{g}})^{2}}{\varGamma(1-i\frac{\tau}{\beta}+\frac{a_{0}}{\beta v_{g}})\varGamma(1+i\frac{\tau}{\beta}+\frac{a_{0}}{\beta v_{g}})}\right)\,,
\end{align}
where $\varGamma$ is the gamma function. In the limit $\frac{a_{0}}{\beta v_{g}}\ll1$ ($\beta=1/T$)
we obtain 
\begin{align}
\mathcal{P}\int_{0}^{\infty}\frac{dk}{k}\left[1-\cos\left(v_{g}k\tau\right)\right]n_{B}(v_{g}k)=\frac{1}{2}\log\left(\frac{\sinh\left(\frac{\pi\tau}{\beta}\right)}{\frac{\pi\tau}{\beta}}\right)\,.\label{eq:I_thermal_Anton}
\end{align}

For obtaining Eq. \eqref{eq:I_thermal_Anton} we used $\varGamma(1-z)\varGamma(1+z)=z\varGamma(z)\varGamma(1-z)=z\pi/\sin(\pi z)$.
Putting together the two contributions we obtain

\begin{align}
I(\tau) & =\frac{1}{2}\log\left[1+\left(\frac{v_{g}\,\tau}{a_{0}}\right)^{2}\right]+\log\left(\frac{\sinh\left(\frac{\pi\tau}{\beta}\right)}{\frac{\pi\tau}{\beta}}\right)\label{eq:integrale1}\\
& \simeq-\log\left[\frac{\frac{\pi a_{0}}{\beta v_{g}}}{\sinh\left(\frac{\pi\tau}{\beta}\right)}\right]\,,
\end{align}
where we approximated $\tau\gg\frac{a_{0}}{v_{g}}=\Lambda_g^{-1}$. Going to large times $\tau\gg\beta$ 
\begin{align}
 I(\tau) \approx \log \left[\frac{\beta \Lambda_g}{2  \pi}\right]+\frac{\pi \tau}{\beta}\,.
\end{align}
Inserting the average into
$Z_{mn}(t)= e^{-\frac{1}{2}\left[\frac{g\left(\lambda_{n}-\lambda_{m}\right)}{\Lambda_g}\right]^{2}\left\langle \left(\varphi_{+}(t)-\varphi_{+}(0)\right)^{2}\right\rangle }$
leads to
\begin{align}
 Z_{mn}(t)\approx\left(\frac{\beta \Lambda_g}{2  \pi} \right)^{-\frac{g}{2}\left[\frac{\left(\lambda_{n}-\lambda_{m}\right)}{\Lambda_g}\right]^{2}}e^{-\Gamma t}\,,
\end{align}
with the decoherence rate $\Gamma$ as given in Eq. \eqref{eq:decoherenza-catastrofe}.

\section{Calculation of  \texorpdfstring{$\tilde{Z}_{12}(t)$}{Z12} and  \texorpdfstring{$\chi_n(\xi,t)$}{Xi}}
\label{App:correlatore}

We evaluate here the expressions in Eqs.~(\ref{decoerenza},\ref{eq:GenFunc} we make use of the fact that in expressions of the form $\langle e^{\pm2i\phi_\pm (\tau)} e^{\pm 2i\phi_\pm} \rangle$ only ``neutral'' configurations of the kind
\begin{align}
f(\tau-\tau') & =\langle e^{2i\varphi_{-}(\tau)}e^{-2i\varphi_{-}(\tau')}\rangle=\langle e^{-2i\varphi_{-}(\tau)}e^{2i\varphi_{-}(\tau')}\rangle\label{correlatore}
\end{align}
do not vanish \cite{giamarchi2004quantum}. Therefore from Eq.\ \eqref{decoerenza}
\begin{align}
\begin{split}\tilde{Z}_{12}(t) & \approx1+\frac{\tilde{\lambda}_{1}\tilde{\lambda}_{2}}{2}\int_{0}^{t}{\rm d}\tau\int_{0}^{t}{\rm d}\tau'f(\tau-\tau')\cos{\left[{\rm{eV}}(\tau-\tau')\right]}\\
 & -\frac{\tilde{\lambda}_{1}^{2}}{2}\int_{0}^{t}{\rm d}\tau\int_{0}^{\tau}{\rm d}\tau'f(\tau-\tau')\cos{\left[{\rm{eV}}(\tau-\tau')\right]}\\
 & -\frac{\tilde{\lambda}_{2}^{2}}{2}\int_{0}^{t}{\rm d}\tau\int_{\tau}^{t}{\rm d}\tau'f(\tau-\tau')\cos{\left[{\rm{eV}}(\tau-\tau')\right]}\,.
\end{split}
\label{eq:Z_mn_cumulant}
\end{align}
Introducing new variables $s=\tau-\tau'$ and $r=(\tau+\tau')/2$,
we can perform the integral over $r$ to obtain 
\begin{align}
\begin{split}\tilde{Z}_{12}(t) & \approx1+\frac{\tilde{\lambda}_{1}}{2}\left(\tilde{\lambda}_{2}-\tilde{\lambda}_{1}\right)\int_{0}^{t}{\rm d}s\,(t-s)f(s)\cos({\rm{eV}}\,s)\\
 & +\frac{\tilde{\lambda}_{2}}{2}\left(\tilde{\lambda}_{1}-\tilde{\lambda}_{2}\right)\int_{-t}^{0}{\rm d}s\,(t-s)f(s)\cos({\rm{eV}}\,s).
\end{split}
\label{eq:Z_mn_Expansion}
\end{align}
We note that 
\begin{align}
\begin{split}f(-s) & =\langle e^{2i\varphi(-s)}e^{-2i\varphi(0)}\rangle=\langle e^{2i\varphi(0)}e^{-2i\varphi(s)}\rangle\\
 & =\langle e^{2i\varphi(s)}e^{-2i\varphi(0)}\rangle^{*}=f(s)^{*},
\end{split}
\end{align}
where in the first equality we make use of the fact that the two-time
correlation function depends only on the time difference. In the long
time limit ($t\gg1/\rm{eV}$), we retain only the dominant contribution for $t\to\infty$,
{\it i.e.} the terms with the integrand $\propto t$ in Eq. (\ref{eq:Z_mn_Expansion}).
Using $f(-s)=f(s)^{*}$ we can re-write the integral in the positive
domain $s>0$ and then replace $f(s)$ by the time-ordered correlator
$f^{T}(s)\equiv\langle\mathcal{T}e^{2i\varphi(s)}e^{-2i\varphi(0)}\rangle$,
which is well known in the literature \cite{giamarchi2004quantum}. We obtain 
\begin{align}\label{eq:tildeZmnApp}
\begin{split}\tilde{Z}_{12}(t) & \approx1-\frac{(\tilde{\lambda}_{2}-\tilde{\lambda}_{1})^{2}}{2}\,t\,\int_{0}^{\infty}{\rm d}s\,{\rm Re}\{f^{T}(s)\}\cos({\rm{eV}}\,s)\\
 & +i\frac{(\tilde{\lambda}_{2}^{2}-\tilde{\lambda}_{1}^{2})}{2}\,t\,\int_{0}^{\infty}{\rm d}s\,{\rm Im}\{f^{T}(s)\}\cos({\rm{eV}}\,s).
\end{split}
\end{align}
Re-exponentiating this expression in the form of Eq. \eqref{eq:Ztilde12}
and disregarding the induced level shift $\tilde{\Delta}$, which leaves
the measurement properties of the device unaffected, leads to Eq.
(\ref{eq:decoerenza-tunneling}) in the main text.

The calculation for the FCS function $W_n(\xi,t)$ proceeds in the same manner, replacing $A_0\rightarrow A_\xi $ in Eq.~\eqref{decoerenza} with $A_\xi$ defined in Eq.~\eqref{eq:Axi} and taking $\lambda_1=\lambda_2=\lambda_n$. We obtain
\begin{align}
W_{n}(\xi,t)\:t=\frac{\tilde{\lambda}_{n}^{2}}{2}\int_{0}^{t}{\rm d}\tau\int_{0}^{t}d\tau'f(\tau-\tau')\:{\rm Re}\left[\left(e^{-i\xi}-1\right)e^{i{\rm{eV}}(\tau-\tau')}\right]\,,
\end{align}
which in the long time limit $t\gg1/\rm{eV}$ leads to Eq.~\eqref{eq:FCS} in the main text.

\section{Calculation of  \texorpdfstring{$\textrm{Re}\left\{ J_{C}\right\} $}{Re(J)} and  \texorpdfstring{$\textrm{Im}\left\{ J_{S}\right\} $}{Im(J)}}
\label{sec:app-E}
In this section we calculate the time integrals $J_{C}$ and $J_{S}$
in Eq. \eqref{eq:decoerenza-tunneling} and \eqref{eq:FCS}. We use
the well known form for the time-ordered correlation function \cite{giamarchi2004quantum}
for positive times 
\begin{align}
f^{T}(s>0) & =\frac{\left( i \frac{\pi a_0}{\beta v_{g}}\right)^{2/g}}{\left(-\sinh\left[\frac{\pi}{\beta}(s-i\,0_{+})\right]\right)^{2/g}}\\
 & =\frac{\left(\frac{\pi a_0}{\beta v_{g}}\right)^{2/g}}{\abs{\sinh^{2}[\frac{\pi}{\beta}t]}^{1/g}}\e^{-i(\pi-0_{+})/g}
\end{align}
Alternatively, this result can be obtained from noting that $f(\tau-\tau')=e^{-\frac{2}{g}I(\tau-\tau' )}e^{2\left[\varphi_-(\tau),\varphi_-(\tau')\right]}$, where $I(s)$ was calculated in App. \ref{integrali}. With this we can evaluate the real part of $J_{C}$, needed for the
decoherence Eq. \eqref{eq:decoerenza-tunneling}. Explicitly,
\begin{eqnarray}
\textrm{Re}\left\{ J_{C}\right\}  & = & \int_{0}^{\infty}{\rm d}s\,\text{Re}\{f^{T}(s)\}\cos({\rm{eV}}\,s)\nonumber \\
 & = & \frac{1}{2}\cos\left(\frac{\pi}{g}\right)\left(\frac{\pi a_0}{\beta v_{g}}\right)^{2/g}\int_{0}^{\infty}{\rm d}s\,\frac{1}{\sinh[\frac{\pi}{\beta}s]^{2/g}}\left(\e^{i{\rm{eV}}s}+\e^{-i{\rm{eV}}s}\right)\nonumber \\
 & = & \frac{1}{2}\cos\left(\frac{\pi}{g}\right)\left(\frac{2\pi a_0}{\beta v_{g}}\right)^{2/g}\frac{\beta}{2\pi}\left(\frac{\varGamma(\frac{1}{g}-i\frac{\beta}{2\pi}{\rm{eV}})\varGamma(1-\frac{2}{g})}{\varGamma(1-\frac{1}{g}-i\frac{\beta}{2\pi}{\rm{eV}})}+\frac{\varGamma(\frac{1}{g}+i\frac{\beta}{2\pi}{\rm{eV}})\varGamma(1-\frac{2}{g})}{\varGamma(1-\frac{1}{g}+i\frac{\beta}{2\pi}{\rm{eV}})}\right)\nonumber \\
 & = & \frac{1}{2}\cos\left(\frac{\pi}{g}\right)\left(\frac{2\pi a_0}{\beta v_{g}}\right)^{2/g}\frac{\beta}{2\pi}\varGamma(1-\frac{2}{g})2\text{Re}\left(\frac{\varGamma(\frac{1}{g}+i\frac{\beta}{2\pi}{\rm{eV}})}{\varGamma(1-\frac{1}{g}+i\frac{\beta}{2\pi}{\rm{eV}})}\right)
\end{eqnarray}
where we dropped the positive infinitesimal $0_{+}$ and used

\begin{align}
\int_{0}^{\infty}{\rm d}s\,\frac{1}{\sinh[\frac{\pi}{\beta}t]^{2/g}}\e^{i\omega t}=2^{2/g}\frac{\beta}{2\pi}B\left(-i\frac{\beta}{2\pi}\omega+\frac{1}{g},1-\frac{2}{g}\right)
\end{align}
with $B(x,y)=\frac{\varGamma(x)\varGamma(y)}{\varGamma(x+y)}$. Using the general
identity of the $\varGamma$-function $\varGamma(x)\varGamma(1-x)=\frac{\pi}{\sin\left(\pi x\right)}$
and some trigonometric identities, we can write this as

\begin{align}
2\cos\left(\frac{\pi}{g}\right)\varGamma(1-\frac{2}{g})\text{Re}\left(\frac{\varGamma(\frac{1}{g}+i\frac{{\rm{eV}}}{2\pi T})}{\varGamma(1-\frac{1}{g}+i\frac{{\rm{eV}}}{2\pi T})}\right)=\frac{\vert\varGamma\left(\frac{1}{g}+i\frac{{\rm{eV}}}{2\pi T}\right)\vert^{2}}{\varGamma\left(\frac{2}{g}\right)}\cosh({\rm{eV}}/2T)\,,
\end{align}
which leads to 
\begin{align}\label{eq:JC}
\textrm{Re}\left\{ J_{C}\right\} =\frac{1}{2}\left(\frac{2\pi a_0}{\beta v_{g}}\right)^{2/g}\frac{\beta}{2\pi}\frac{\vert\varGamma\left(\frac{1}{g}+i\frac{{\rm{eV}}}{2\pi T}\right)\vert^{2}}{\varGamma\left(\frac{2}{g}\right)}\cosh({\rm{eV}}/2T)\,,
\end{align}
in accordance with the results in Ref. \onlinecite{Furusaki1998b}.
From here the form of the decoherence rate $\tilde{\Gamma}$ 
Eq. \eqref{eq:GammaTilde} directly follows.

Similarily, $\textrm{Im}\left\{ J_{S}\right\} $ can be calculated to be
\begin{eqnarray*}
\text{Im}\{J_{S}\} & = & \int_{0}^{\infty}{\rm d}s\,\,\text{Im}\left\{ f^{T}(s)\sin({\rm{eV}}\,s)\right\} \\
 & = & -\left(\frac{\pi a_0}{\beta v_{g}}\right)^{2/g}\sin\left(\frac{\pi}{g}\right)\int_{0}^{\infty}{\rm d}s\,\frac{1}{\abs{\sinh^{2}[\frac{\pi}{\beta}t]}^{1/g}}\frac{1}{2i}\left(\e^{i{\rm{eV}}s}-\e^{-i{\rm{eV}}s}\right)\\
 & = & -\frac{1}{2i}\sin\left(\frac{\pi}{g}\right)\left(\frac{2\pi a_0}{\beta v_{g}}\right)^{2/g}\frac{\beta}{2\pi}\left(\frac{\varGamma(\frac{1}{g}-i\frac{\beta}{2\pi}{\rm{eV}})\varGamma(1-\frac{2}{g})}{\varGamma(1-\frac{1}{g}-i\frac{\beta}{2\pi}{\rm{eV}})}-\frac{\varGamma(\frac{1}{g}+i\frac{\beta}{2\pi}{\rm{eV}})\varGamma(1-\frac{2}{g})}{\varGamma(1-\frac{1}{g}+i\frac{\beta}{2\pi}{\rm{eV}})}\right)\\
 & = & \sin\left(\frac{\pi}{g}\right)\left(\frac{2\pi a_0}{\beta v_{g}}\right)^{2/g}\frac{\beta}{2\pi}\varGamma(1-\frac{2}{g})\text{Im}\left(\frac{\varGamma(\frac{1}{g}+i\frac{\beta}{2\pi}{\rm{eV}})}{\varGamma(1-\frac{1}{g}+i\frac{\beta}{2\pi}{\rm{eV}})}\right)
\end{eqnarray*}
Using again $\varGamma(x)\varGamma(1-x)=\frac{\pi}{\sin\left(\pi x\right)}$
and some trigonometric identities, we can write this as

\begin{align}
-2\sin\left(\frac{\pi}{g}\right)\varGamma(1-\frac{2}{g})\text{Im}\left(\frac{\varGamma(\frac{1}{g}+i\frac{{\rm{eV}}}{2\pi T})}{\varGamma(1-\frac{1}{g}+i\frac{{\rm{eV}}}{2\pi T})}\right)=\frac{\vert\varGamma\left(\frac{1}{g}+i\frac{{\rm{eV}}}{2\pi T}\right)\vert^{2}}{\varGamma\left(\frac{2}{g}\right)}\sinh({\rm{eV}}/2T)\,,
\end{align}
which sets the form of $\text{Im}\{J_{S}\}$

\begin{align}
\text{Im}\{J_{S}\}=\frac{1}{2}\left(\frac{2\pi a_0}{\beta v_{g}}\right)^{2/g}\frac{\beta}{2\pi}\frac{\vert\varGamma\left(\frac{1}{g}+i\frac{{\rm{eV}}}{2\pi T}\right)\vert^{2}}{\varGamma\left(\frac{2}{g}\right)}\sinh({\rm{eV}}/2T)\,.
\end{align}
\end{widetext}

\end{document}